\documentclass[10pt,english,prl,showpacs,superscriptaddress]{revtex4}
\usepackage[latin9]{inputenc}
\usepackage{textcomp}
\usepackage{amsmath}
\usepackage{amssymb}
\usepackage{graphicx}

\makeatletter
\@ifundefined{textcolor}{}
{%
 \definecolor{BLACK}{gray}{0}
 \definecolor{WHITE}{gray}{1}
 \definecolor{RED}{rgb}{1,0,0}
 \definecolor{GREEN}{rgb}{0,1,0}
 \definecolor{BLUE}{rgb}{0,0,1}
 \definecolor{CYAN}{cmyk}{1,0,0,0}
 \definecolor{MAGENTA}{cmyk}{0,1,0,0}
 \definecolor{YELLOW}{cmyk}{0,0,1,0}
}

\usepackage{epstopdf}
\usepackage{color}
\makeatother

\usepackage{babel}
\begin{document}

\title{Cavity solitons in vertical-cavity surface-emitting lasers}

\author{A. G. Vladimirov$^{1}$, A. Pimenov$^{1}$, S. V. Gurevich$^{2}$, K. Panajotov$^{3,4}$, E. Averlant$^{3,5}$, and M. Tlidi$^{5}$}

\address{Weierstrass Institute, Mohrenstrasse 39, D-10117 Berlin, Germany $^{1}$\\
Institute for Theoretical Physics, University of M{\"u}nster, Wilhelm-Klemm-Str.9, D-48149, M{\"{u}}nster, Germany$^{2}$\\
Department of Applied Physics and Photonics (IR-TONA), Vrije Unversiteit Brussels, Pleinlaan 2, B-1050 Brussels, Belgium$^{3}$ \\
Institute of Solid State Physics, 72 Tzarigradsko Chaussee Blvd., 1784 Sofia, Bulgaria$^{4}$\\
Universit\'{e} Libre de Bruxelles (U.L.B.), Facult{\'{e}} des Sciences,  CP. 231, Campus Plaine, B-1050 Bruxelles, Belgium$^{5}$
}

\begin{abstract}
We investigate a control of the motion of localized structures of light by means of delay feedback in the transverse section of a broad area nonlinear optical system. The delayed feedback is found to induce a spontaneous motion of a solitary localized structure that is stationary and stable in the absence of feedback. We focus our analysis on an experimentally relevant system namely the Vertical-Cavity Surface-Emitting Laser (VCSEL). In the absence of the delay feedback we present experimental evidence of stationary localized structures in a  80  $\mu$m aperture VCSEL. The spontaneous formation of localized structures takes place above the lasing threshold and under optical injection. Then, we consider the effect of the time-delayed optical feedback and investigate analytically the role of the phase of the feedback  and the carrier lifetime on the self-mobility properties of the localized structures. We show that these two parameters affect strongly the space time dynamics of two-dimensional localized structures. We derive an analytical formula for the threshold associated with drift instability of localized structures and a normal form equation describing the slow time evolution of the speed of the moving structure.
\end{abstract}


\maketitle\section{Introduction}

Transverse localized structures often called cavity solitons were observed experimentally in Vertical-Cavity Surface-Emitting Laser (VCSEL) \cite{Taranenko2000,Barland}. Due to the maturity of the semiconductor technology and the possible applications of localized structures of light in all-optical delay lines \cite{Pedaci08} and logic gates \cite{Jacobo2012}, these structures in VCSELs have been a subject of an active research in the field of nonlinear optics. Moreover, the fast response time of VCSELs makes them attractive devices for potential applications in all-optical control of light. Localized structures appear as solitary peaks or dips on homogeneous background of the field emitted by a nonlinear microresonator with a high Fresnel number. Here, we report experimental evidence of spontaneous formation of localized structures in a 80 $\mu$ m diameter VCSEL biased above the lasing threshold and under optical injection. Such localized structures are bistable with the injected beam power and the VCSEL current.

Various mechanisms have proven to be responsible for the generation of localized structures (LS) in VCSELs: coherent optical injection (holding beam, i.e. the part of the optical injection that is used to ensure bistability of the system) in combination with a narrow writing beam (the part of optical injection which is used to locally make the system evolve from the lower branch of the bistability curve to the higher branch and vice versa)\cite{Barland}, frequency selective feedback with \cite{Tanguy08} or without \cite{Radwell} a writing beam, saturable absorption \cite{Genevet}, amplifiers \cite{Menesguen}, optically injected lasers \cite{Larionova,Hachair2006}, ``spatial translational coupling'' introduced in \cite{Haudin,delCampo}, and others (see \cite{Barbay} for a review).
LS can start to move spontaneously in the laser transverse section in the presence of saturable absorption \cite{Fedorov}. In particular, in the case when the pump beam has a circular shape LS move along the boundary on a circular trajectory  \cite{PratiEPJD}.  It has been shown  that they can undergo a spontaneous motion under the thermal effects \cite{Spinelli,Scroggie}.
Delayed feedback control is a well documented technique that has been applied to various spatially extended systems  in optics, hydrodynamics, chemistry, and biology. It has been demonstrated recently that a simple feedback loop provides a robust and a controllable mechanism for the motion of localized structures (LS) and localized patterns \cite{Tlidi09,Panajotov10,Tlidi12,Gurevich13,Tlidi-sonnino,Gurevich13a,Pimenov}. These works demonstrated that when the product of the delay time and the feedback rate exceeds some threshold value, localized structures start to move in an arbitrary direction in the transverse section of the device. In these studies, the analysis was restricted to the specific case of nascent optical bistability described by the real Swift-Hohenberg equation with a real feedback term. More recently, analytical study supported by numerical simulations revealed the role of the phase of the delayed feedback and the carrier lifetime on the motion of cavity solitons in a broad-area VCSEL structure, driven by a coherent externally injected beam.  We show that for certain values of the feedback phase LS can be destabilized via a drift bifurcation leading to a spontaneous motion of a solitary two-dimensional LS \cite{Tlidi09}. We demonstrate also that the slower is the carrier decay rate in the semiconductor medium, the higher is the threshold associated with the motion of cavity solitons.

The paper is organized as follows: In Sec. 2, we discuss the experimental observation of stationary LS in medium size VCELS. In Sec. 3, we introduce model equations. In Sec. 4, we investigate the drift instability induced by delay feedback. We conclude in Sec. 5.
\section{Experimental observation of stationary LS in medium size VCELS}

In recent years, a considerable amount of experimental work has been realized on stationary localized structures in Vertical-Cavity Surface-Emitting Lasers (VCSELs).  They were observed in very broad (aperture $d > 100 \mu m$) \cite{Barland}, broad  ($d \sim 80 \mu$m) \cite{Averlant_oe14}, and medium ($d \sim 40 \mu m$) \cite{Hachair_pra09} size VCSELs.  Here we present experimental results obtained with bottom-emitting InGaAs multiple quantum well VCSEL with $d=80 \mu m$ and threshold current of 42.5 mA at 20$^\circ$C. The holding (injection) beam is provided by a commercial tunable semiconductor laser (Sacher Lasertechnik TEC100-0960-60 External Cavity Diode Laser), isolated from the rest of the setup by an optical isolator (OFR IO5-TiS2-HP). The long-term electrical and temperature stability of this laser are less than 20 mA RMS and 0.05 $^\circ$C, respectively. A half-wave plate is used to adjust the linear polarization of the holding beam to be the same as the one of the VCSEL. The injection beam power is tuned using a variable optical density filter. The detuning between the master laser and the VCSEL is defined as $\theta=\nu_{inj} -\nu_{slave}$, where $\nu_{inj}$ is the frequency of the injection beam, and $\nu_{slave}$ the frequency of the strongest peak in the spectrum of the stand-alone VCSEL. It is experimentally tuned by changing the wavelength of the injection beam. The beam waist $d_{inj}$ is defined as the diameter of the smallest circle in the plane of propagation of the injection beam containing half of the beam power when it encounters the VCSEL. The power of the source is monitored by a Newport 818-SL photodiode connected to a Newport 2832-C powermeter. The near field is recorded by imaging it on a CCD camera.

\begin{figure}[!h]
\centering\includegraphics[width=4.5in]{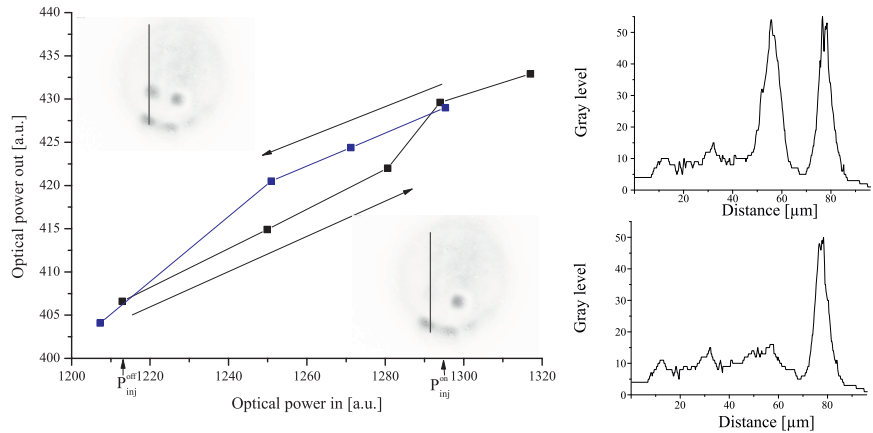}
\caption{Total output power as a function of injection power displaying bistability when a new LS appears. The insets show the near field images of on the upper and the lower branch of the hysteresis curve. The two figures on the right show one dimensional scans along the vertical lines indicated in the near field images.}
\label{fig_1}
\end{figure}

An example of stationary LSs is presented in the Fig.~\ref{fig_1} illustrating the process of spontaneous creation and annihilation of two-dimensional localized structures. These experimental measurements have been performed when the VCSEL operated in an injection locked regime. The injection beam waist and the detuning are fixed to  $d_{inj} = 50 \mu$m and $\theta$= -146 GHz.  When increasing the injected beam power, a new LS appears at $P_{inj}=P_{inj}^{on}$ as shown in the insets of Fig.~\ref{fig_1}. This results in a slight jump of the total output power as shown in light-versus-current characteristics. The process of switching-off is realized when decreasing the injection power: the recently created LS persists until $P_{inj}=P_{inj}^{off}$ with $P_{inj}^{off}<P_{inj}^{on}$, i.e. a hysteresis region exists with an additional LS either turned on or off. The two figures on the right show one dimensional scans along the vertical lines indicated in the near field images.

\section{Model equations}
The mean field model describing the space-time evolution of the electric field envelope $E$ and  the  carrier density $N$ in a VCSEL subjected to optical injection is given by the following set of dimensionless partial differential equations
\begin{eqnarray}
\frac{\partial E}{\partial t} &=& -\left(\mu+i\theta\right)E + 2C(1-i\alpha)(N-1)E  \label{eq:dEdt} \\
&+& E_{i} -\eta e^{i\varphi} E(t-\tau)+ i\nabla^{2} E\,, \nonumber \\
\frac{\partial N}{ \partial t} &=& -\gamma \left[ N -I + (N-1)\left|
E\right| ^{2} - d\nabla^{2} N\right]\,. \label{eq:dNdt}
\end{eqnarray}
Here the parameter $\alpha$ describes the linewidth enhancement factor, $\mu$ and $\theta$ are the cavity decay rate and the cavity detuning parameter, respectively. Below we will assume $\eta$ to be small enough, so that we can neglect the dependence of the parameters $\mu$ and $\theta$ on $\varphi$. The parameter $E_{i}$ is the amplitude of the injected field,  $C$ is the bistability parameter, $\gamma$ is the carrier decay rate, $I$ is the injection current, and $d$ is the carrier diffusion coefficient. The diffraction of light and the diffusion of the carrier density are described by the terms $i\nabla^{2}E$ and $d\nabla^{2}N$, respectively, where $\nabla^{2}$ is the Laplace operator acting in the transverse plane $(x,\,y)$. Below we consider the case when the laser is subjected to coherent delayed feedback from an external mirror. To minimize the effect of diffraction on the feedback field we assume that the external cavity is self-imaging \cite{Tlidi09}. The feedback is characterized by the delay time $\tau=2L_{ext}/c$, the feedback rate $\eta\ge 0$, and phase $\varphi$, where $L_{ext}$ is the external cavity length, and $c$ is the speed of light. The link between dimensionless and physical parameters is provided in \cite{Panajotov10}. Using the expression for the feedback rate $\eta=\frac{r^{1/2}(1-R)}{R^{1/2}\tau_{in}}$ given in \cite{Tartwijk}, where $r$ ($R$) is the power reflectivity of the feedback (VCSEL top) mirror and $\tau_{in}$ is the VCSEL cavity round trip time, we see that the necessary condition for the appearance of the soliton drift instability $\eta\tau>1$ \cite{Tlidi09} can be rewritten in the form $r>\frac{R \tau_{in}^2}{(1-R)^{2}\tau^2}$. In particular, for $R=0.3$ and $\tau=20\tau_{in}$ the latter inequality becomes $r>1.5\cdot 10^{-3}$.

\section{Drift instability threshold}

\begin{figure}[!h]
\centering\includegraphics[width=4.5in]{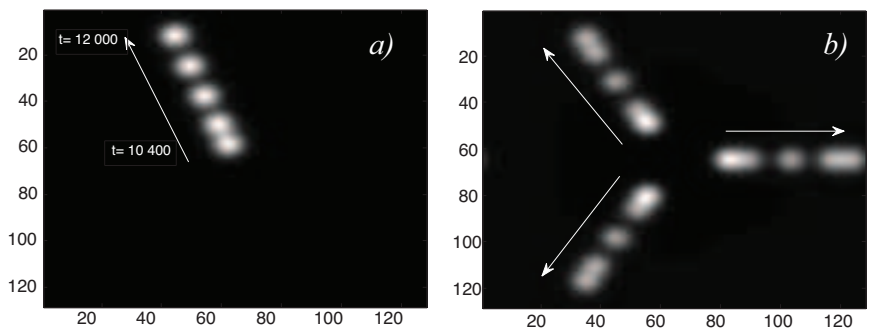}
\caption{Field intensity illustrating a moving a single (a) and three (b) peak  LS. Parameter values are C=0.45, $\theta=-2.$,  $\alpha=5.$, $\mu=1.$, 
Feedback parameters are $\eta=0.135$, $\tau=100$, $\varphi=0.5$. Maxima are plain white.}
\label{fig_numerics}
\end{figure}

When the delayed feedback is absent, $\eta=0$,  Eqs.~(\ref{eq:dEdt}) and (\ref{eq:dNdt}) are transformed into the well-known mean field model \cite{Spinelli_pra98}, which supports stable stationary patterns and LSs \cite{Barland,Hachair04,Hachair05,Hachair_pra09}. It was demonstrated recently that when the feedback rate $\eta$ exceeds a certain threshold value, which is inversely proportional to the delay time $\tau$, LS starts to move in the transverse direction \cite{Tlidi09}.  Example of moving two-dimensional LS are shown in Fig.~\ref{fig_numerics}. The single and the three moving peaks are obtained from numerical simulations of Eqs.~(\ref{eq:dEdt}) and (\ref{eq:dNdt}). The boundary condition are periodic in both transverse dimensions.

\begin{figure}[!h]
\centering\includegraphics[width=5.0in]{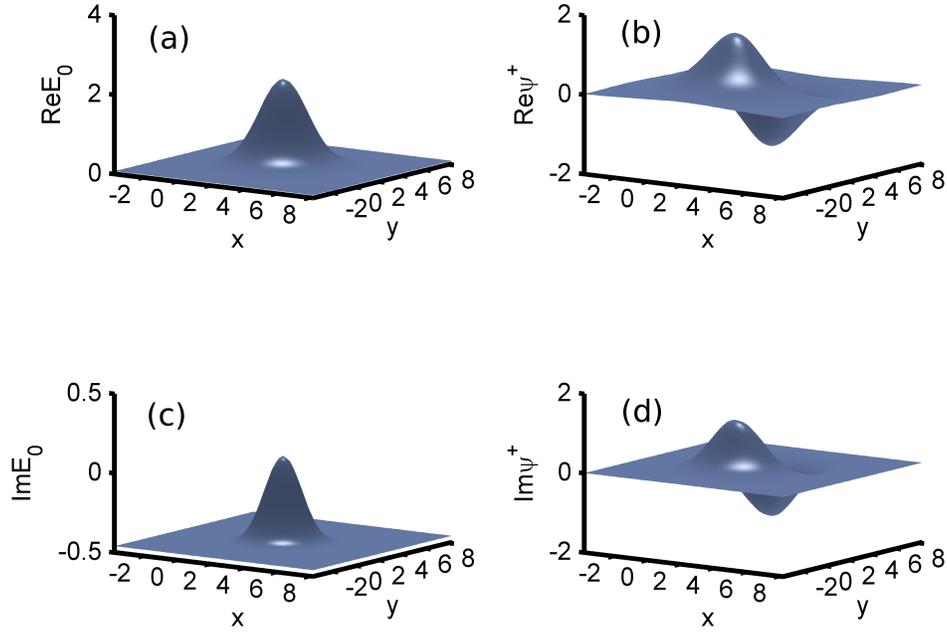}
\caption{Left panels: real and imaginary parts of the stationary soliton profile, $X_0=\Re E_0$ (a)  $Y_0=\Im E_0$ (c). Right panels: real and imaginary parts of the neutral mode of the adjoint operator $L^{\dagger}$, $\psi^{\dagger}_1=\Re \psi^{\dagger}$ (b) and $\psi^{\dagger}_2=\Im \psi^{\dagger}$ (d). Parameters values: $\mu = 1.0$, $\theta=-2.0$, $C=0.45$, $\alpha=5.0$, $\gamma=0.05$, $\tau=100$, $d = 0.052$, $E_i = 0.8$, $I = 2$.
}
\label{soliton}
\end{figure}

 In the case when the system is transversely isotropic, the velocity of the LS motion has an arbitrary direction. The self-induced motion of the LS is associated with a pitchfork bifurcation where the stationary LS looses stability and a branch of stable LSs uniformly moving with the velocity $v=|{\mathbf v}|$ bifurcates from the stationary LS branch. The bifurcation point can be obtained from the first order expansion of the uniformly moving LS in power series of the small velocity $v$.  Close to the pitchfork bifurcation point this expansion reads:   
\begin{eqnarray}
E(x-v t,y) = E_0(x -v t,y) + v E_1(x -v t,y) + ...\\
N(x-v t,y) = N_0(x -v t,y) + v N_1(x -v t,y)+...,
\label{expansion}
\end{eqnarray}
where without the loss of generality we assume that the LS moves along the $x$-axis on the $(x,y)$-plane. Here $E_0(x,y)=X_0(x,y) + i Y_0(x,y)$ and $N_0(x,y)$ describes the stationary axially symmetric LS profile, which corresponds to the time-independent solution of Eqs.~(\ref{eq:dEdt}) and (\ref{eq:dNdt}) with $\tau=0$. Although formally this solution depends on the feedback parameters $\eta$ and $\varphi$ we neglect this dependence assuming that the feedback rate is sufficiently small, $\eta\ll 1$. Substituting this expansion into Eqs.~(\ref{eq:dEdt}) and (\ref{eq:dNdt}) and collecting the first order terms in small parameter $v$ we obtain:

\begin{eqnarray}
L \left(\begin{array}{c}\Re E_{1}\\ \Im E_{1}\\N_1\end{array}\right) = \left(\begin{array}{c}\Re[\partial_x E_0(1 - \eta \tau e^{i\varphi})]\\ \Im[\partial_x E_0(1 - \eta \tau e^{i\varphi})] \\\gamma^{-1}\partial_x N_0\end{array}\right)\label{eq:1storder}
\end{eqnarray}
where the linear operator $L$ is given by

$$
L = \left(\begin{array}{ccc}\mu - 2 C(N_0-1) & \nabla^2-\theta- 2 C\,\alpha\, (N_0-1) & -2C(X_{0} + \alpha Y_{0})\\
-\nabla^2+\theta+ 2 C\,\alpha\, (N_0-1)  &\mu-2C(N_0-1) & -2 C (Y_{0} - \alpha X_{0})\\
 2(N_0-1) X_{0} & 2(N_0-1) Y_{0} &-d \nabla^2+1+|E_0|^2 \end{array}\right).
$$
By applying the solvability condition to the right hand side of Eq.~(3), we obtain the drift instability threshold
\begin{equation}
\eta\tau =\frac{1+\gamma^{-1}(b/c) }{ \sqrt{1+(a/c)^2}\cos[\varphi+\arctan{(a/c)}]}
\label{thresholdlimit}
\end{equation}
with
\begin{equation}
a= \langle \psi_1^{\dagger}, \psi_2 \rangle-\langle \psi_2^{\dagger}, \psi_1\rangle,\quad
b=\langle \psi_3^{\dagger}, \psi_3 \rangle,\quad
c=\langle \psi_1^{\dagger},\, \psi_1\rangle + \langle \psi_2^{\dagger},\,\psi_2\rangle.\label{coeff}
\end{equation}
Here \begin{equation}
{\psi}=\left(\psi_1,\,\psi_2,\,\psi_3\right)^T=\partial_x\left(X_0,\,Y_0,\,N_0\right)^T \label{nm}
\end{equation} 
is a translational neutral mode of the operator $L$, $L\psi=0$, while  ${\psi}^{\dagger}=\left(\psi_1^{\dagger},\,\psi_2^{\dagger},\,\psi_3^{\dagger}\right)^T$  is the corresponding solution of the homogeneous adjoint problem $L^\dagger \psi^{\dagger} = 0$. The scalar product $\langle \cdot \rangle$ is defined as  $\langle \psi_j^{\dagger},\, \psi_k\rangle= \int_{-\infty}^{+\infty}\psi_{j}^\dagger\psi_{k}\,dxdy$. To estimate the coefficients $a$ and $b$ we have calculated the function $\psi^{\dagger}$ numerically using the relaxation method in two transverse dimensions, $(x,y)$. The results of these calculations are shown in Fig.~\ref{soliton} together with the axially symmetric profile $E_0$ of the stationary LS. It is seen from this figure that similarly to the neutral mode $\psi$ defined by (\ref{nm}) the neutral mode $\psi^{\dagger}$ of the adjoint operator $L^{\dagger}$ is an even function of the coordinate $y$ and an odd function of the coordinate $x$, which is parallel to the LS direction of motion.

\begin{figure}[!h]
\centering\includegraphics[width=4.0in]{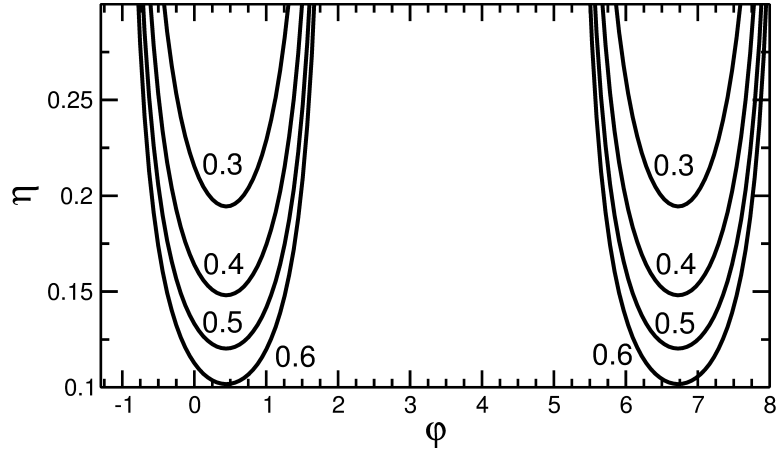}
\caption{Critical value of the feedback rate $\eta$ corresponding to the drift bifurcation vs feedback phase $\varphi$ calculated for different values of the carrier relaxation rate $\gamma$. The values of the parameter $\gamma$ are shown in the figure. Other parameters are the same as in Fig.~\ref{soliton}.
}
\label{gamma}
\end{figure}

The dependence of the critical feedback rate $\eta$ corresponding to the drift instability threshold defined by Eq.~(\ref{thresholdlimit}) on the feedback phase $\varphi$ and carrier relaxation rate $\gamma$ is illustrated by Fig.~\ref{gamma}. In this figure the curves labeled by different numbers correspond to different values of $\gamma$. Considering the fact that the feedback in Eq.~(\ref{eq:dEdt}) is introduced with the minus sign, we see that the drift instability takes place only for those feedback phases when the interference between the cavity field and the feedback field is destructive, i.e. when $\cos$ function in the denominator of the right hand side of Eq.~(\ref{thresholdlimit}) is positive. On the contrary, when this interference is constructive the feedback has a stabilizing effect on the LS. Furthermore, the slower is the carrier relaxation rate, the higher is the drift instability threshold.
Since the stationary LS solution does not depend on the carrier relaxation rate $\gamma$, the coefficients $a$ and $b$ in the threshold condition (\ref{thresholdlimit}) are also independent of $\gamma$. Therefore, (\ref{thresholdlimit}) gives an explicit dependence of the threshold feedback rate on the carrier relaxation rate. In particular, in the limit of very fast carrier response, $\gamma \gg 1$, and zero feedback phase, $\varphi=0$, we recover from (\ref{thresholdlimit}) the threshold condition $\eta\tau = 1$ which was obtained earlier for the LS drift instability induced by a delayed feedback in the real Swift-Hohenberg equation \cite{Tlidi09}. Note that at $\gamma\to\infty$, $a\neq 0$, and $\varphi=-\arctan{a}$ the critical feedback rate appears to be smaller than that obtained for the real Swift-Hohenberg equation, $\eta\tau={(1+a^2)}^{-1/2}<1$.

\begin{figure}[!h]
\centering\includegraphics[width=4.0in]{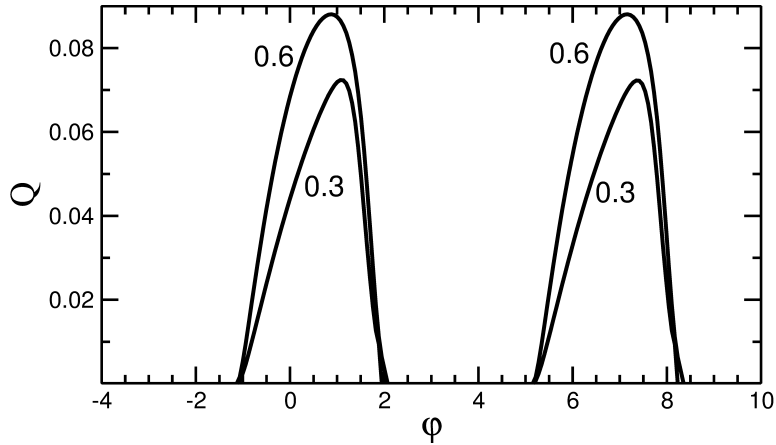}
\caption{Coefficient $Q$ describing the grows rate of the LS velocity with the square root of the deviation from the critical feedback rate. The values of the parameter $\gamma$ are shown in the figure. Other parameters are the same as in Fig.~\ref{soliton}.
}
\label{q}
\end{figure}

As it was demonstrated above, the bifurcation threshold responsible for self-induced drift of LS in the VCSEL transverse section is obtained by expanding the slowly moving localized solution in the small velocity $v$, substituting this expansion into the model equations (\ref{eq:dEdt}),(\ref{eq:dNdt}), and equating the first order terms in $v$. In order to describe the slow evolution of the LS velocity slightly above the bifurcation threshold one needs to perform a similar procedure with $E=E_0(x-x_0(t),y)+\sum_{k=1}^3\epsilon^kE_k(x-x_0(t),y,t)+...$ and $N=N_0(x-x_0(t),y)+\sum_{k=1}^3\epsilon^kN_k(x-x_0(t),y,t)+...$, where $dx/dt=v(t)={\cal O}(\epsilon)$, $dv/dt={\cal O}(\epsilon^3)$ and $\epsilon$ is a small parameter characterizing the distance from the bifurcation point. Then, omitting detailed calculations, in the third order in $\epsilon$, we obtain the normal form equation for the LS velocity:
\begin{equation}
\frac{p}{2}\frac{d v}{dt}=v(\delta\eta q  - \eta\tau^2 r v^2), \label{nf}
\end{equation}
where $\delta\eta$ is the deviation of the feedback rate from the bifurcation point. The coefficients $q$, $p$, and $r$ are given by $q=a\sin{\varphi}+c\cos{\varphi}$, $p=q+b$, and $r=f\sin{\varphi}+g\cos{\varphi}+{\cal O}(\tau^{-1})$, respectively. Here $a$, $b$, and $c$ are defined by Eq.~(\ref{coeff}) and $f=\langle \psi_1^{\dagger}, \partial_{xxx} Y_0\rangle-\langle \psi_2^{\dagger}, \partial_{xxx} X_0\rangle$, $h=\langle \psi_3^{\dagger}, \partial_{xxx} N_0 \rangle$, $g=\langle \psi_1^{\dagger},\, \partial_{xxx} X_0\rangle + \langle \psi_2^{\dagger},\,\partial_{xxx} Y_0\rangle$. The stationary LS velocity above the drift instability threshold is obtained by calculating the nontrivial steady state of Eq.~(\ref{nf}), $v=\sqrt{\delta\eta} Q$, where the coefficient $Q=(1/\tau)\sqrt{q/(r\eta)}$ determines how fast the LS velocity increases with the square root of the deviation from the critical feedback rate. The dependence of this coefficient on the feedback phase is illustrated by Fig. \ref{q}.

\section{Conclusion}

To conclude, we have studied experimentally the formation of transverse LS in a medium size bottom-emitting InGaAs multiple quantum well VCSEL operated in an injection locked regime. Creation and annihilation of a single localized structure have been demonstrated by changing the injection beam power. Theoretically we have analyzed the effect of delayed feedback from an external mirror on the stability of transverse LS in a broad area VCSEL.  We have shown that depending on the phase of the feedback it can have either destabilizing or stabilizing effect in the LS. In particular, when the interference between the LS field and the feedback field is destructive, the LS can be destabilized via a pitchfork bifurcation, where a branch of uniformly moving LS bifurcates from this the stationary one. We have calculated analytically the threshold value of the feedback rate corresponding to this bifurcation and demonstrated that the faster the carrier relaxation rate in the semiconductor medium, the lower is the threshold of the spontaneous drift instability induced by the feedback. Finally, we have derived the normal for equation (\ref{nf}) governing the slow dynamics of the LS velocity.

\section*{Acknowledgment}

A.G.V. and A.P. acknowledge the support from SFB 787 of the DFG. A.G.V. acknowledges the support of the EU FP7 ITN PROPHET and E.T.S. Walton Visitors Award of the Science Foundation Ireland. M.T. received support from the Fonds National de la Recherche Scientifique (Belgium). This research was supported in part by the Interuniversity Attraction Poles program of the Belgian Science Policy Office under Grant No. IAP P7-35.


\end{document}